\documentclass{revtex4-1}
\usepackage{amsmath,amssymb,color,graphics}
\usepackage{bm}
\usepackage[
colorlinks=true,backref,pagebackref]{hyperref} 

\definecolor{dark-green}{rgb}{0,0.7,0}
\definecolor{dark-blue}{rgb}{0,0.2,0.5}
\definecolor{med-blue}{rgb}{0,0.7,1}
\definecolor{mblue}{rgb}{0,0.2,1}
\definecolor{cnc}{rgb}{0.8,0,0}
\definecolor{light-red}{rgb}{1,0.8,0.8}
\definecolor{dark-yellow}{rgb}{1,0.8,0}
\definecolor{light-blue}{rgb}{0.8,0.9,1}
\definecolor{verylight-blue}{rgb}{0.93,0.95,1}
\definecolor{light-yellow}{rgb}{1,0.9,0.8}
\definecolor{grey}{gray}{0.88}

\def\a{\alpha}
\def\b{\beta}
\def\c{\gamma}
\def\r{\rho}
\def\d{\delta}
\def\g{\gamma}
\def\m{\mu}
\def\n{\nu}

\def\vt{\vartheta}

\def\ep{\varepsilon}
\def\vt{\vartheta}
\def\vp{\varphi}

\usepackage{amsfonts}

\begin{document}
\title{Premetric equivalent of general relativity: Teleparallelism}

\author{Yakov \surname{Itin}} \email{itin@math.huji.ac.il}
\affiliation{Institute of Mathematics, The Hebrew University of Jerusalem\\
  Jerusalem College of Technology, Jerusalem 91160, Israel}

\author{Friedrich W. \surname{Hehl}}
\email{hehl@thp.uni-koeln.de}\homepage{http://www.thp.uni-koeln.de/gravitation/mitarbeiter/hehl.html}
\affiliation{Institute for Theoretical Physics, University of Cologne,
  50923 Cologne, Germany\\
  Department of Physics and Astronomy, University of Missouri,
  Columbia, MO 65211, USA}

\author{Yuri N. \surname{Obukhov}}
\email{obukhov@ibrae.ac.ru} 
\affiliation{Theoretical Physics Laboratory, Nuclear Safety Institute,
  Russian Academy of Sciences, B. Tulskaya 52, 115191 Moscow, Russia}

\date{09 Apr 2017}

\begin{abstract}
  In general relativity (GR), the metric tensor of spacetime is
  essential since it represents the gravitational potential. In other
  gauge theories (such as electromagnetism), the so-called premetric
  approach succeeds in separating the purely topological field
  equation from the metric-dependent constitutive law. We show here
  that GR allows for a premetric formulation, too. For this purpose,
  we apply the teleparallel approach of gravity, which represents GR
  as a gauge theory based on the translation group. We formulate the
  metric-free topological field equation and a general linear
  constitutive law between the basic field variables. The requirement
  of local Lorentz invariance turns the model into a full equivalent
  of GR.  Our approach opens a way for a natural extension of GR to
  diverse geometrical structures of spacetime.  {\it file
    teleGRprd12.tex}
\end{abstract}
\maketitle

\tableofcontents
\section{Introduction} 

{\it The premetric formalism} is an alternative representation of a
classical field theory in which the field equations are formulated
without the spacetime metric.  Only the constitutive relations between
the basic field variables, excitation $H$ and field strength $F$, can
involve the metric of the underlying manifold. This idea can be traced
back to the early 1920s where it appears in the publications of
Kottler \cite{Kottler:1922a}, \cite{Kottler:1922b}. Various
applications of this construction to the formal structure of
electrodynamics were worked out by Post \cite{Post:1962}. The
premetric formalism was studied intensively in the book
\cite{Birkbook}. For an account of the recent developments in this
area, see {our} review \cite{Hehl:2016glb}.

{One advantage of the premetric formalism is that the
  validity of a certain premetric model can be extended to a more
  general spacetime geometry.} The premetric construction works pretty
well in Maxwell's classical electrodynamics. In this case, all basic
ingredients, such as the field equations, the conserved quantities of
electric charge and of magnetic flux, and the Lorentz force expression
are presented in a metric-free form. Only the constitutive relation
between the excitation and the field strength are formulated with the
use of the metric tensor. {And this relation can be
  straightforwardly extended to a local and linear relation thereby
  getting rid of the metric altogether.} Let us briefly recall the
various outputs of this approach:
\begin{itemize}
\item Natural extension of standard electrodynamics by axion,
  skewon, and dilaton fields;
\item metric-free dispersion relation for electromagnetic waves in
  a medium with general linear response behavior;
\item metric-free Green tensor (photon propagator);
\item metric-free jump conditions that include boundary conditions
  between two media, initial Cauchy and wave-front
  conditions;
\item derivation of the metric from the local and linear constitutive
  relation by prohibiting birefringence in electromagnetic wave
  propagation;
\item natural account of Lorentz violation models.
\end{itemize}

Although Kottler's premetric program works well in Newtonian gravity
\cite{Kottler:1922a} and even in a flat gravitomagnetism model
\cite{Hehl:2016glb}, it seems to be completely unacceptable in general
relativity (GR). This is due to the well-known fact that Einstein's
theory is essentially based on a pseudo-Riemann geometry with the
metric tensor as its primary variable. Nevertheless, in this paper, we
will show that a premetric construction of GR is possible if one turns
to its teleparallel reformulation.

The organization of the paper is as follows: In Section 2, we
construct a teleparallel model for the coframe field. It is a
vector-valued analog of electromagnetic theory with a { well-defined
  gravitational energy-momentum current and a Lorentz-type force
  density.} The general local linear constitutive law between the
coframe excitation and the coframe field strength is defined by the
use of a constitutive tensor density of 6th rank.  In Section 3, we
consider the coframe model on a pseudo-Riemannian manifold. This
restriction naturally requires the localization of the group of
coframe transformations.  Moreover, when the constitutive tensor
density is assumed to be constructed from the metric, the model turns
out to be {fully equivalent to GR.} Section 4 is devoted to the
Lorentz force density as an interaction term in the equation of
{motion for a particle.} We construct a metric-free equation for a
congruence of trajectories with a constitutive law between the
momentum {covector and the velocity vector.} Its restriction to the
metric manifold yields {a} geodesic curve in the {gravitational} case
and a trajectory of a charge in an exterior field in the
electromagnetic case.  In the concluding section, we discuss the main
properties of our construction and propose some {possible} extensions
of standard GR.  In the Appendix we provide some technical
calculations.

\section{Premetric electrodynamics and its coframe analog in gravity}

As was shown in \cite{Birkbook}, classical electrodynamics can be
expressed in a premetric way. In this section, we briefly recall the
basic electromagnetic quantities and construct their coframe analogs.

Our key assumption is that a gauge field model of gravity must be
based on a conserved current, here on the macroscopic (``bosonic'')
energy-momentum current of matter, see Blagojevi\'c et al.\
\cite{Blagojevic:2013xpa}. This is in analogy to the electric current
that serves as a basis of electromagnetic theory. We use a
{co}vector-valued 3-form as a representation of the material
energy-momentum current and construct a vector-valued
field-theoretical model. It represents a {\it vector-valued} analog of
the electromagnetic theory. Recall that the latter is expressed in
terms of ordinary, scalar-valued differential forms. Although at this
stage, our construction appears to be only formal, its justification
is based on its relation to the energy-momentum conservation
law. Incidentally, the existence of an additional independent
conserved 2-form, which is untwisted, is naturally related to the
definition of a special coframe field on the manifold.

\subsection{Geometric structure}

Let us consider a differential manifold ${\cal M}$ endowed with a coframe
field $\vt^\a $. The 1-forms $\vt^\a $, with $\a=0,1,2,3$, are assumed
to be linearly independent at each point of ${\cal M}$.  At this
stage, we postulate that all equations are invariant under {\it rigid
  linear transformations} of the coframe $\vt^\a$. The transformed
coframe $\vt^{\a'}$ then becomes
\begin{equation}\label{geo4}
\vt^{\a'}= L_\a{}^{\a'} \vt^\a\,,\quad L_\a{}^{\a'}=\text{const}\,,
\end{equation}
with a constant invertible matrix $L_\a{}^{\a'}\in GL(4,\mathbb{R})$. 

The coframe and its exterior products (taken in increasing order)
generate the bases
\begin{equation}\label{geo1}
  \vt^{\a}\,, \quad \vt^{\a\b}:=\vt^{\a}\wedge\vt^{\b}\,, \quad 
  \vt^{\a\b\g}:=\vt^{\a}\wedge\vt^{\b}\wedge\vt^{\g}\,, \quad
  \vt^{\a\b\g\d}  :=\vt^{\a}\wedge\vt^{\b}\wedge\vt^{\g}\wedge\vt^{\d}
\end{equation}
of the spaces of untwisted differential forms of the order 1, 2, 3,
and 4, respectively.  Under the transformation (\ref{geo4}), the basis
forms (\ref{geo1}) transform as tensors.

In order to express the {\it twisted forms}, we need the volume
element (a non-negative measure) defined on $\cal M$. Relative to the basis
$\vt^\a$, it is defined as a twisted scalar-valued 4-form
\begin{equation}\label{geo2}
{\rm vol} = {\frac 1{4!}}\,\varepsilon_{\a\b\g\d}\,\vt^\a\wedge
\vt^\b\wedge\vt^\g\wedge\vt^\d\otimes s\,, 
\end{equation}
Here $\varepsilon_{\a\b\g\d}$ is the Levi-Civita permutation symbol
\cite{Levi} that is normalized to $\varepsilon_{0123}=1$, while $s$ is a
section of the orientation line bundle. In \cite{Bott-Tu},
Eq.(\ref{geo2}) is represented symbolically as the absolute value of
the untwisted 4-form.  Under the transformation of the coframe
(\ref{geo4}), the volume element (\ref{geo2}) transforms {according
  to} the law
 \begin{equation}\label{geo4x}
{\rm vol}\longrightarrow |\det L|\,{\rm vol}\,,
\end{equation}  
with $\det L$ as the Jacobian of the coframe transformation. Thus, the
volume element (\ref{geo2}) remains positive for all admissible
coframes.

The frame field $e_a$ is uniquely defined as the inverse of the coframe, 
\begin{equation}\label{geo0'}
e_\a\rfloor \vt^\b=\vt^\b(e_\a)=\d^\b_\a\,.
\end{equation}
Under the coframe transformation (\ref{geo4}), the frame obeys the
transformation law
\begin{equation}\label{geo0}
e_{\a} \longrightarrow  e_{\a'}=(L^{-1})_{\a'}{}^{\,\a}e_\a\,,
\end{equation}
with $(L^{-1})_{\a'}{}^{\a}L_\a{}^{\b'}=\d_{\a'}^{\b'}$.

With these definitions at hand, the sets
\begin{equation}\label{geo3}
  {\rm vol}\,, \quad\epsilon_\a=e_\a\rfloor{\rm vol}\,, \quad
{\epsilon_{\a\b}=e_\b\rfloor\epsilon_\a\,,}\quad
    \epsilon_{\a\b\g}=e_\g\rfloor\epsilon_{\a\b}\,,\quad
\epsilon_{\a\b\g\d}=e_\d\rfloor\epsilon_{\a\b\g}\,,
\end{equation}
with the indices taken in increasing order, serve as basis forms for
the spaces of twisted 4-forms, 3-forms, 2-forms, 1-forms, and 0-forms,
respectively. These basis forms transform with an additional factor
$|\det L|$. All forms in (\ref{geo3}) are totally antisymmetric. 
It is worthwhile to note that the Levi-Civita permutation symbol
$\varepsilon_{\a\b\g\d}$ is an untwisted tensor density, and one can 
check that the values of its components do not change under the frame
transformation. In contrast, the 0-form $\epsilon_{\a\b\g\d}$ is a twisted
density, which means that its components, so to say, are sensitive
to the orientation: they either remain the same or change 
their sign when the frame transformation preserves orientation or
changes orientation, respectively. In technical terms, the behavior 
of $\epsilon_{\a\b\g\d}$ depends on the sign of determinant $\det L$. 
This explains a different notation for $\epsilon_{\a\b\g\d}$ and 
$\varepsilon_{\a\b\g\d}$. From the definitions (\ref{geo3}) we can
straightforwardly check the identity
\begin{equation}\label{vev}
\vartheta^\alpha\wedge\epsilon_\beta = \delta^\alpha_\beta\,{\rm vol}.
\end{equation}

Since at the first stage, we allow only {\it global} (rigid)
transformations of the coframe, the exterior derivatives of the basis
forms (\ref{geo1}) and (\ref{geo3}) transform as tensors.  Hence, one
does not need here an exterior {\it covariant} derivative of the forms. 
Subsequently, we will discuss how the symmetry transformation
(\ref{geo4}) can be generalized to the case of {a point dependent}
$L_\a{}^{\a'}(x)$.

\subsection{Excitation} {\bf Electromagnetism:} In electromagnetism,
the inhomogeneous field equation can be treated as a result of the
electric charge conservation law. In order to describe, in a given
spatial volume, the electric charge with a prescribed sign, we must
use the twisted 3-form $J$ of the electric current. Its expression in
a twisted basis reads
\begin{equation}\label{el-kot1a}
J=J^\a\epsilon_\a\,.
\end{equation}
Under the coframe transformations (\ref{geo4}), we have
$\epsilon_\a\longrightarrow  \epsilon_{\a'}=|\det L|
(L^{-1})_{\a'}{}^\a\epsilon_\a$, and the components of the 3-form
$J$ transform as
\begin{equation}\label{el-kot1b}
J^\a\longrightarrow  J^{\a'}=(\det L)^{-1}L_\a{}^{\a'} J^\a\,,
\end{equation}
or
\begin{equation}\label{el-kot1c}
J\longrightarrow  \frac{|\det L|} {\det L}\,J\,.
\end{equation}
The 3-form $J$ remains the same under orientation preserving
transformations, while picking up an additional sign under
transformations which reverse the orientation of the coframe. This
additional sign compensates the change of the orientation of the
integration domain.  Consequently, the integral $\int_{\Omega_3} J$
(in particular, the total charge for a closed spatial domain
$\Omega_3$) is invariant under the coframe transformations.
 
The law of electric charge conservation in integral and differential
forms is given by
\begin{equation}\label{el-kot1}
  \int_{\partial\Omega_4} J=0\quad\text{and}\quad dJ=0\,,
\end{equation}
respectively. Locally, the latter relation is equivalent to the
inhomogeneous Maxwell equation
\begin{equation}\label{el-kot2}
d{H}=J\,,
\end{equation}
where $ H$  is the twisted 2-form of the {\it electromagnetic excitation}.  
In the $\vt^{\a\b}$ and $\epsilon_{\a\b}$ bases, it reads
\begin{equation}\label{el-kot2a}
  H=\frac 12 {H}_{\a\b}\vt^{\a\b}=\frac 12 \check{H}^{\a\b}\epsilon_{\a\b}\,,
  \quad\text{with}\quad \check{H}^{\a\b}=\frac 12\epsilon^{\a\b\g\d}H_{\g\d}\,. 
\end{equation}
By construction, ${H}_{\a\b}$ is a covariant twisted tensor, whereas 
$\check{H}^{\a\b}$ is an untwisted contravariant tensor density. \\

\noindent {\bf Gravity:} Similarly to this electrodynamics
construction, we start our gravity model with a conservation law, now
with energy-momentum conservation.  In the canonical formalism,
the standard energy-momentum tensor is replaced by the energy-momentum
current $\Sigma_\a$, a twisted covector-valued 3-form. We decompose it
with respect to the 3-forms $\epsilon_\b$,
\begin{equation}\label{gr-kot0}
\Sigma_\a=\Sigma_\a{}^\b\,\epsilon_\b\,.
\end{equation}
This is an object of 16 independent components, see
\cite{Hehl:1994ue}. Symmetry may only be imposed by the use of a
metric tensor.

Taking into account (\ref{geo4}), with constant $L_\b{}^\a$, the
conservation law for the energy-momentum current can be expressed as
\begin{equation}\label{gr-kot1}
\int_{\partial\Omega_4} \Sigma_\a=0\,, \qquad d\,\Sigma_\a=0\,.
\end{equation}
Using the standard differential-geometric facts, we can solve
Eq.(\ref{gr-kot1}) in a small topologically good region as
\begin{equation}\label{gr-kot2}
d{H}_\a=\Sigma_\a\,.
\end{equation}
In this way, we define (up to a total derivative) the twisted
covector-valued {\it gravitational excitation} 2-form
\begin{equation}\label{gr-kot2a}
{H}_\a=\frac 12{H}_{\b\g}{}_\a\,\vt^{\b\g} =\frac 12
\check{H}^{\b\g}{}_\a\,\epsilon_{\b\g}\,.
\end{equation}
It is of decisive importance to recognize that there is a fundamental
difference to the electromagnetic case (\ref{el-kot2}). The
electromagnetic field does not carry electric charge (the ``photon''
is electrically neutral), the gravitational field, however, carries
energy-momentum of its own. Hence the right-hand side of
(\ref{gr-kot2}) reads
$\Sigma_\a=\,^{\text{(m)}}\Sigma_\a+\,^{(\vartheta)}\Sigma_\a$. Here
(m) denotes matter and $(\vartheta)$ the coframe field, and we assume
additivity of the {corresponding energy-momenta.}

\subsection{Field strength}

{\bf Electromagnetism:} In electrodynamics, the untwisted field
strength 2-form
 \begin{equation}\label{el-kot3a}
 F=\frac 12 F_{\a\b}\vt^{\a\b}
\end{equation}
 satisfies the equations 
 \begin{equation}\label{el-kot3}
\int_{\partial\Omega_3} F=0\,, \qquad d{ F}=0\,.
\end{equation}
The homogeneous Maxwell equation $d{F}=0$ is an expression of the
conservation of the magnetic flux. The electromagnetic field strength
$F$ is determined operationally via the Lorentz force density, which
acts on the electric current. We will discuss this below.  The
solution of Eq.(\ref{el-kot3}) can be expressed in terms of the {\it
  electromagnetic potential} $A$,
 \begin{equation}\label{el-kot3b}
dA=F\,.
\end{equation}
In the coframe basis, this untwisted 1-form reads 
 \begin{equation}\label{el-kot3c}
A=A_\a\vt^\a\,.
\end{equation}
It is defined up to the addition of a total derivative $A\longrightarrow  A +d\vp$.

\vspace{0.4cm}

\noindent {\bf Gravity:} In analogy to the field strength $F$ of
the electromagnetic theory, we introduce the {\it gravitational field
  strength} $F^\a$. It is an untwisted vector-valued 2-form that
satisfies the equation
 \begin{equation}\label{gr-kot3}
d{ F}^\a=0\,.
\end{equation}
The solution of this equation can be locally represented as 
 \begin{equation}\label{gr-kot4}
{ F}^\a=d\theta^\a\,.
\end{equation}
The set of four 1-forms $\theta^\a$ is the analog of the
electromagnetic potential $A$.  We assume now that the potentials
$\theta^\a$ are linearly independent. It always can be reached due to
the gauge invariance of Eq.(\ref{gr-kot3}). Indeed, we can redefine
$\theta^\a\longrightarrow  \theta^\a+df^a$, with four arbitrary scalar functions
$f^\a$.

We identify the reference coframe $\vartheta^\a$ with the dynamical
coframe $\theta^\a$ and rewrite Eq.(\ref{gr-kot4}) as
 \begin{equation}\label{gr-kot4a}
{ F}^\a=d\vartheta^\a\,.
\end{equation}
Thus, we can consider the covector-valued forms $\Sigma_\a, { H}_\a$
and the vector-valued $F^\a$ to be related to this special basis.  In
particular, we expand the untwisted form $F^\a$ relative to the
untwisted basis $\vartheta^{\b\g}$ as follows:
 \begin{equation}\label{gr-kot5}
F^\a=\frac 12 F_{\b\g}{}^\a\vartheta^{\b\g}\,.
\end{equation}

\subsection{Lorentz force}

{\bf Electromagnetism:} The force acting on electrically
  charged matter is described by a twisted covector-valued 4-form
  $f_\a$. Being a top-order form, it can be represented as a
  vector-valued scalar ${\frak f}_\a$ multiplied by the volume form
  $f_\a={\frak f}_\a{\rm vol}$. In electrodynamics, see
  \cite{Birkbook} and also \cite{Bamberg,Thirring,Bladel}, 
the Lorentz force is given by
\begin{equation}\label{el-lor1}
f_\a=\left(e_\a\rfloor F\right) \wedge J\,.
\end{equation}
Readers can refer to \cite{Birkbook,Bamberg,Thirring,Bladel} for technical
details explaining how one can compute the electric power and the total force 
of electromagnetic field acting on the matter by taking an appropriate integral 
of the Lorentz force density (\ref{el-lor1}).  

Expanding the current with respect to the 3-form basis, 
\begin{equation}\label{el-lor1a}
  J = J^\a \epsilon_\a\,,
\end{equation}
and making use of (\ref{vev}), we recast the Lorentz force (\ref{el-lor1}) into
\begin{equation}\label{el-lor1c}
f_\a=\left(J^\b F_{\a\b}\right) {\rm vol}\,.
\end{equation}
The first factor represents the standard expression of the Lorentz force 
density
\begin{equation}\label{el-lor1cx}
{\frak f}_\a = J^\b F_{\a\b}\,.
\end{equation}
By construction, $J^\a$ is an untwisted vector density, and accordingly 
${\frak f}_\a$ is an untwisted covector density. For a point particle, both 
the current density and the force density are proportional to a delta-function 
\cite{deFelice:1990}. 

\vspace{0.4cm}

\noindent {\bf Gravity:} Analogously to electromagnetism, we describe 
the Lorentz force for the coframe field by the 4-form
\begin{equation}\label{cof-lor1b}
f_\a=\left(e_\a\rfloor F^\b\right) \wedge {}^{\text{(m)}}\Sigma_\b\,.
\end{equation}
Expanding the energy-momentum current with respect to the 3-form basis, 
\begin{equation}\label{cof-lor1c}
  {}^{\text{(m)}}\Sigma_\a = \Sigma_\a{}^\b \epsilon_\b\,,
\end{equation}
we introduce an untwisted energy-momentum tensor density $\Sigma_\a{}^\b$
of massive matter.
Substituting the representation (\ref{gr-kot5}) into (\ref{cof-lor1b})
and using (\ref{cof-lor1c}), we obtain the gravitational Lorentz force
\begin{equation}\label{cof-lor1d}
f_\a=\left(\Sigma_\g{}^\b\! F_{\a\b}{}^\g \right) {\rm vol}\,.
\end{equation}
The first factor on the right-hand side of (\ref{cof-lor1d})
represents the covector of the {\it gravitational} Lorentz force density
\begin{equation}\label{el-lor1cy}
{\frak f}_\a = \Sigma_\g{}^\b\! F_{\a\b}{}^\g \,.
\end{equation}
A comparison between (\ref{el-lor1cy}) and (\ref{el-lor1cx}) shows the
deep analogy between gravity and electromagnetism.

\subsection{Energy-momentum current of gravity}

{\bf Electromagnetism:} The energy-momentum current of the
electromagnetic field, see \cite{Birkbook}, is a covector-valued
3-form represented by
\begin{equation}\label{el-energy} 
{}^{\rm(em)}\Sigma_\a={\frac 12}  \left[F\wedge 
(e_\a\rfloor H)-H\wedge  (e_\a\rfloor F)\right]\,.
\end{equation}
If the twisted electromagnetic Lagrangian 4-form
 \begin{equation}\label{el-action}
{}^{\rm(em)}\Lambda:=-\frac 12 F\wedge H\,
\end{equation}
can be specified, we can alternatively put it into the form
\begin{eqnarray}\label{el-energy1} {}^{\rm(em)}\Sigma_\a=e_\a\rfloor
  {}^{\rm(em)}\Lambda+F\wedge (e_\a\rfloor H) =-e_\a\rfloor
  {}^{\rm(em)}\Lambda-H\wedge (e_\a\rfloor F)\,.
\end{eqnarray}
Using $F=dA$, we can rederive the field equation $dH=J$ and the current
(\ref{el-energy}) from the Lagrangian (\ref{el-action}).

One can straightforwardly verify the balance law \cite{Birkbook} 
\begin{equation}
d\,{}^{\rm(em)}\Sigma_\a = f_\a + X_\a,\label{balanceE}
\end{equation}
where $f_\a$ is the Lorentz force (\ref{el-lor1}) and $X_\a = -\,{\frac 12}
(F\wedge {\cal L}_\a H - H\wedge {\cal L}_\a F)$ describes an additional
force determined by the constitutive law. Here ${\cal L}_\a$ denotes the
Lie derivative along vector fields $e_\a$. 

\vspace{0.4cm}

\noindent{\bf Gravity:}
Similar to the electromagnetic case, we postulate the energy-momentum
current of the coframe field as
 \begin{equation}\label{cof-energy}
{}^{(\vartheta)}\Sigma_\a=\frac 12 \left[F^\b\wedge (e_\a\rfloor  H_\b)-H_\b
\wedge (e_\a\rfloor F^\b)\right]\,.
\end{equation}
We can also introduce the {Lagrange} 4-form for the coframe field,
\begin{equation}\label{cof-action} {}^{(\vartheta)}\Lambda=-\frac 12
  F^\a\wedge H_\a\,.
\end{equation}
Then we can write its energy-momentum current in {a form}
similar to (\ref{el-energy1}),
 \begin{eqnarray}\label{cof-energy1}
{}^{(\vartheta)}\Sigma_\a=e_\a\rfloor
   {}^{(\vartheta)}\Lambda+F^\b\wedge (e_\a\rfloor H_\b)
   =-e_\a\rfloor {}^{(\vartheta)}\Lambda-H_\b\wedge (e_\a\rfloor F^\b)\,.
\end{eqnarray}
Analogously to (\ref{balanceE}), one finds the balance law  
\begin{equation}
d\,{}^{(\vartheta)}\Sigma_\a = f_\a + {}^{(\vartheta)}\!X_\a,\label{balanceG}
\end{equation}
where $f_\a$ is gravitational Lorentz force (\ref{cof-lor1b}) and ${}^{(\vartheta)}
\!X_\a = -\,{\frac 12}(F^\b\wedge {\cal L}_\a H_\b - H_\b\wedge {\cal L}_\a F^\b)$ 
is an additional force to be determined by the corresponding constitutive law.

\subsection{Constitutive relation}

In order to complete the field-theoretical models of electromagnetism
and gravity, a {\it constitutive relation} between the basic
variables, namely {between} excitation $H$ and field strength $F$
should be introduced.

\vspace{0.4cm}

\noindent{\bf Electromagnetism:} The system of the premetric field
equations for electromagnetism (\ref{el-kot2}) and (\ref{el-kot3})
involves 8 equations for 12 independent variables, the components of
the 2-forms $H$ and $F$. This system is undetermined and has to be
supplemented by an additional relation between the basic variables. In
solid state electromagnetism, such relation can be of a rather
complicated form. However, even the simplest case of a linear
constitutive relation has a wide range of applications.

Using the expansions 
\begin{equation}\label{el-const'}
 H = {\frac 12}\check{H}^{\a\b}\epsilon_{\a\b}\,, \qquad  F=\frac 12 F_{\a\b}\vt^{\a\b}\,,
\end{equation}
we postulate the most general local linear constitutive relation in
the form of
 \begin{equation}\label{el-const}
\check{H}^{\a\b}=\frac 12\chi^{\a\b\g\d}F_{\g\d}\,.
\end{equation}
Due to this definition, the constitutive tensor density $\chi$
satisfies the symmetry relations
 \begin{equation}\label{el-const-sym}
\chi^{\a\b\g\d}=-\chi^{\b\a\g\d}=-\chi^{\a\b\d\g}\,.
\end{equation}

\vspace{0.4cm}

\noindent{\bf Gravity:} Similarly, our coframe system must be endowed
with the constitutive relation between $F^\a$ and ${H}_\a$. We assume
this relation to be linear and local. In analogy to electromagnetism,
we use the expansions
  \begin{equation}\label{gr-kot9a}
    H_\a=\frac 12\check{H}^{\b\g}{}_{\a}\epsilon_{\b\g}\,,\qquad 
F^\a=\frac 12  F_{\b\g}{}^\a\,\vartheta^{\b\g}\,.
\end{equation}
We postulate the most general local and linear constitutive relation
in the form of
 \begin{equation}\label{gr-kot9}
   \check{H}^{\b\g}{}_{\a}=\frac 12
   \chi^{\b\g}{}_{\a}{}^{\nu\rho}{}_\mu F_{\nu\rho}{}^\mu\,.
\end{equation}
Here $\chi^{\b\g}{}_{\a}{}^{\nu\rho}{}_\mu $ is the constitutive
tensor density that obeys the symmetries
 \begin{equation}\label{gr-kot10}
\chi^{\b\g}{}_{\a}{}^{\nu\rho}{}_\mu =- \chi^{\g\b}{}_{\a}{}^{\nu\rho}{}_\mu 
= - \chi^{\b\g}{}_{\a}{}^{\rho\nu}{}_\mu \,.
\end{equation}

\subsection{Lagrange formalism} 

In this section, we apply the Lagrange formalism to derive the
statements proposed above. In this way, we are able to justify the
coframe model that was postulated in the previous section only by
analogy.

\vspace{0.4cm}

\noindent{\bf Electromagnetism:} Although the electromagnetic case is
well-known, it is instructive to recall the variational
procedure. This construction turns out to be completely metric-free.
As only restriction, we will use an additional symmetry relation of
the constitutive tensor density, namely
\begin{equation}\label{el-lag0}
\chi^{\a\b\c\d}=\chi^{\c\d\a\b}\,.
\end{equation}
In term of the irreducible decomposition \cite{Birkbook}, it means
that the skewon part is assumed to be forbidden and the constitutive
tensor density is left with only 21 independent components; then and
only then a Lagrange formalism is possible.

We start with the Lagrange 4-form
\begin{equation}\label{el-lag1}
  \Lambda=-\frac 12 F\wedge H(F) +A\wedge J=\left(-\frac 12F_{\a\b}
    \check{H}^{\a\b}(F_{\g\delta})+A_\a J^\a\right)\,{\rm vol}\,.
\end{equation}
The variation of this Lagrangian takes the form
 \begin{equation}\label{el-lag2}
   \d \Lambda=-\frac 12 \left(\d F\wedge H+F\wedge \d H\right)+\d A\wedge J\,.
\end{equation}
In the case of the linear constitutive relation with the symmetry
(\ref{el-lag0}), the first two terms on the right-hand side of
Eq.(\ref{el-lag2}) are equal to one another.  Indeed, using the component
representation, we have
\begin{eqnarray}\label{el-lag3}
  F\wedge \d H&=&-\frac 12\left(F_{\a\b}\d\check{H}^{\a\b}\right)\!{\rm vol}
=-\frac 14\left(F_{\a\b}\chi^{\a\b\c\d}\d F_{\c\d}\right)\!{\rm vol}\nonumber\\ 
&=&-\frac 12\left(\check{H}^{\c\d}\d F_{\c\d}\right)\!{\rm vol}= \d F\wedge  H\,.
\end{eqnarray}
Consequently, Eq.(\ref{el-lag2}) takes the form
 \begin{eqnarray}\label{el-lag4}
   \d \Lambda=-d(\d A)\wedge H+\d A\wedge J=-d(\d A\wedge H)-\d A\wedge(dH-J)\,.
\end{eqnarray}
In order to derive the field equation from this expression, we remove,
as usual, the total derivative term and require $\d \Lambda$ to be
zero for arbitrary variations of the potential. Then we obtain the
inhomogeneous Maxwell equation and the electric charge conservation
law as straightforward consequences,
\begin{equation}\label{el-lag5}
dH=J\,, \qquad dJ=0\,.
\end{equation}
Let us now study relation (\ref{el-lag4}) on shell, i.e, we assume
that the inhomogeneous Maxwell equation (\ref{el-lag5}) is already
satisfied. Then we are left with
 \begin{equation}\label{el-lag6}
\d \Lambda=-d(\d A\wedge H)\,.
\end{equation}
For variations induced by frame transformations, we use
$\d_\a \Lambda$ instead of $\d \Lambda$ and $\d_\a A$ instead of
$\d A$. These variations are generated by the Lie derivatives relative
to the frame vectors, $\d_\a ={\cal L}_{e_\a}$. Thus, we have
\begin{eqnarray}\label{el-lag7a}
\d_\a \Lambda&=&{\cal L}_{e_\a}\Lambda=d(e_\a\rfloor \Lambda)\,,\\ 
\label{el-lag7b}
\d_\a A&=&{\cal L}_{e_\a}A=d(e_\a\rfloor A)+e_\a\rfloor dA\,.
\end{eqnarray}
Substituting into (\ref{el-lag6}), we obtain a conservation law
 \begin{equation}\label{el-lag8}
d\, {}^{\text{(em)}\!}\Sigma_\a=0\,,
\end{equation}
where
 \begin{equation}\label{el-lag9}
{}^{\text{(em)}\!}\Sigma_\a=\big[e_\a\rfloor \Lambda+\left(e_\a\rfloor
    F\big)  \wedge H\right]-(e_\a\rfloor A)\wedge J\,.
\end{equation}
On the right-hand side of this equation, we recognize the
energy-momentum of the electromagnetic field and the interaction term.


\vspace{0.4cm}

\noindent{\bf Gravity:}
Consider a Lagrangian of a system that includes the coframe field and
a matter field
\begin{equation}\label{lag1}
\Lambda=\frac 12 F^\a\wedge H_\a+{}^{\rm (m)}\Lambda\,.
\end{equation}
Using (\ref{gr-kot9a}), we rewrite it as
\begin{equation}\label{lag2}
  \Lambda=\frac 12 \left(F_{\b\c}{}^\a \check{H}^{\b\c}{}_\a\right)\!{\rm vol}
  +{}^{\rm (m)}\Lambda\,.
\end{equation}
Variation of this Lagrangian  reads (see Appendix)
\begin{equation}\label{lag3}
  \d\Lambda=-d(\d\vt^\a\wedge H_\a) -\d\vt^\a\wedge( d H_\a
  - {}^{(\vartheta)}\Sigma_\a-{}^{\rm (m)}\Sigma_\a)\,,
\end{equation}
where the energy-momentum current of the coframe field is specified by
\begin{equation}\label{lag3x} 
{}^{(\vartheta)}\Sigma_\a=e_\a\rfloor
  \Lambda+F^\b\wedge (e_\a\rfloor H_\b)\,.
\end{equation}
The matter energy-momentum current ${}\,^{\rm (m)\!}\Sigma_\a$ is defined
via the relation
\begin{equation}\label{lag4}
\d\,{}^{\rm (m)\!}\Lambda=\d\vt^\a\wedge{}^{\rm (m)\!}\Sigma_\a\,.
\end{equation}
For variations of the coframe that vanish on the boundary, we are left
with the field equation
\begin{equation}\label{lag5}
d H_\a= \Sigma_\a\,,
\end{equation}
where the total energy-momentum current is given as a sum of the
coframe current (\ref{lag3x}) and the matter current defined in (\ref
{lag4})
\begin{equation}\label{lag6}
  \Sigma_\a={}^{(\vartheta)}\Sigma_\a+{}^{\rm (m)}\Sigma_\a\,.
\end{equation}
{Note} that the conservation law for this quantity, 
$d \Sigma_\a=0$,
follows straightforwardly from field equation (\ref {lag5}).
 
\subsection{Premetric electromagnetism-gravity correspondence}

We can now summarize the analogy between the premetric coframe model
of gravity and the standard electromagnetic theory in Table~\ref{table}.

\begin{table}
\caption{\label{table}Premetric electromagnetism-gravity analogy.}
\begin{tabular}{|c|c|c|}
  \hline 
  Objects and Laws & Electromagnetism & Gravity  \\
  \hline \hline 
  Source current& $J$& $\Sigma_\a$\\
  \hline 
  Conserved source current  &$dJ=0$&$d\Sigma_\a=0$\\
  \hline 
  Excitation &$H$&$H_\a$\\
  \hline 
  Inhomogeneous field equation &$dH=J$&$dH_\a
  ={}^{(\vartheta)}\Sigma_\a+{}^{\text{(m)}}\Sigma_\a$\\
  \hline
  Field strength&$F$&$F^\a$\\
  \hline
  Homogeneous field equation&$dF=0$&$dF^\a=0$\\
  \hline
  Potential&$A$&$\vt^\a$\\
  \hline
  Potential equation&$dA=F$&$d\vt^\a=F^\a$\\
  \hline
  Lorentz force &$f_\a=\left(e_\a\rfloor F\right) \wedge J$&$f_\a
=\left(e_\a\rfloor F^\b\right) \wedge {}^{\text{(m)}}\Sigma_\b$\\
  \hline
  Energy-momentum current&$\Sigma_\a=e_\a\rfloor \Lambda+F\wedge (e_\a\rfloor H)$
&${}^{(\vartheta)}\Sigma_\a=e_\a\rfloor \Lambda+F^\b\wedge (e_\a\rfloor H_\b)$\\
  \hline
  Lagrangian&$\Lambda=- (1/2) F\wedge H$&$\Lambda=- (1/2) F^\a\wedge H_\a$\\
  \hline 
  Constitutive tensor &$\chi^{\a\b\g\d}$&$\chi^{\b\g}{}_{\a}{}^{\nu\rho}{}_\mu$\\
  \hline
\end{tabular}
\end{table}

\section{Field-theoretical models on metric manifolds }

So far, all the ingredients in the electromagnetic as well as in the
coframe model are premetric. Indeed, the metric is not involved in
these formalisms at all. We will now consider these models on a
manifold endowed with a pseudo-Riemannian metric. In the
electromagnetic case, this structure allows us to describe vacuum
electrodynamics.  For the coframe field, we are able to reinstate
standard GR in the context of a premetric formalism.

\subsection{Coframe field and metric} 

We consider a manifold ${\cal M}$ endowed with a smooth metric $g$ and
restrict the coframe field $\vartheta^a$ to be orthonormal relative to
this metric. Thus, the metric on our manifold can be expressed as
 \begin{equation}\label{gr-kot7}
g = g_{\a\b}\,\vartheta^\a\otimes \vartheta^\b\,,
\end{equation}
where $g_{\a\b}={\rm diag}(+1,-1,-1,-1)$ is the Minkowski metric. In
other words, we restrict ourselves to the subgroup $ O(1,3)$ of the
orthogonal transformations of the coframe:
 \begin{equation}\label{gr-kot7y}
\vt^\a\longrightarrow  \vt^{\a'}= L_\a{}^{\a'} \vt^\a\,.
\end{equation}
Then the metric satisfies the relation
 \begin{equation}\label{gr-kot7z}
g_{\a'\b'} L_\a{}^{\a'}  L_\b{}^{\b'}  = g_{\a\b}\,,
\end{equation}
and hence $(\det L)^2 = 1$. 
We observe that the metric in (\ref{gr-kot7}) is invariant
under a wider class of transformations that depend on a point
$x\in{\cal M}$ with $L_\a{}^{\a'}(x)$, that is, we have {\it local
  coframe transformations}.

We can develop the coframe and the frame fields, respectively, in
terms of local coordinates $\{x^i\}$ as follows:
 \begin{equation}\label{gr-kot7a}
\vartheta^\a=\vartheta_i{}^\a dx^i\,,\qquad e_\a=e^i{}_\a\partial_i\,.
\end{equation}
In these holonomic coordinates, the components of the metric tensor
read
 \begin{equation}\label{gr-kot8}
   g_{ij}=g_{\a\b}\,\vartheta_i{}^\a\vartheta_j{}^\b\,,\qquad g^{ij}
   =g^{\a\b} e^i{}_\a e^j{}_\b\,.
\end{equation}
The volume element (\ref{geo2}) takes now the form
 \begin{equation}\label{gr-kot8'}
{\rm vol}=\sqrt{-g}\,d^4x=|\det \vt_i{}^\a| d^4x\,,
\end{equation}
where $g=\det \left(g_{ij}\right) = -(\det \vartheta_i{}^\alpha)^2$. We recognize 
in this standard expression the twisted 4-form as defined in (\ref{geo2}).

It is worthwhile to note that $\check{H}^{\a\b}$ and $\check{H}^{\a\b}{}_\g$ are
true tensors under the restriction to the orthogonal group.

\subsection{Vacuum electrodynamics} 

Standard Maxwell-Lorentz electrodynamics is recovered in the
premetric framework provided the constitutive tensor is expressed in
terms of the Minkowski metric as follows:
 \begin{equation}\label{el-vac}
\chi^{\a\b\g\d} = {\frac 12}\,\lambda_0
\left(g^{\a\g}g^{\b\d}-g^{\a\d}g^{\b\g}\right)\,.
\end{equation}
Here $g$ is the determinant of the metric and
$\lambda_0=\sqrt{\varepsilon_0/\mu_0}$ denotes the vacuum
impedance. In `the International System of Units' (SI), its value is
$\lambda_0=1/(377\,\Omega)$. If we only allow the metric $g^{\a\b}$ to
enter the constitutive tensor (\ref{el-vac}) as variable, then, due to
the symmetries (\ref{el-const-sym}) of $\chi^{\a\b\g\d}$, the
construction of (\ref{el-vac}) is well determined.

We expand the  field strength 2-form in a coordinate basis 
\begin{equation}\label{el-vac1}
  F=\frac 12 F_{ij}dx^i\wedge dx^j\,
\end{equation}
and derive from $dF=0$ the homogeneous Maxwell equation in
{its} standard form in tensor calculus:
 \begin{equation}\label{el-vac2}
 \epsilon^{ijkl}{\partial_j}F_{kl}=0\,.
\end{equation}
If the constitutive tensor (\ref{el-vac}) is used, also the
inhomogeneous field equation $dH=J$ can be rewritten in the standard
tensor notation,
 \begin{equation}\label{el-vac3}
{\partial_j\left(\sqrt{-g}F^{ij}\right)}=\sqrt{-g}J^i\,.
\end{equation}
This results in the conservation law of the electric current,
 \begin{equation}\label{el-vac3x}
\partial_i\left(\sqrt{-g}J^i\right)=0\,.
\end{equation}
The Lorentz force in a {coframe} basis reads
 \begin{equation}\label{el-vac4}
f_\a=e_\a{}^iF_{ik}\sqrt{-g}\,J^kd^4x\,.
\end{equation}
The scalar factor of this 4-form presents the ordinary expression of
the Lorentz force density covector
 \begin{equation}\label{el-vac5}
{\frak f}_i=F_{ik}J^k\,.
\end{equation}

\subsection{Constitutive tensor density of the coframe}

We turn now to the gravitational model.  We require the 6th rank
constitutive tensor density to be expressed in terms of the metric tensor
$g_{\a\b}$ as variable alone. Due to the symmetries listed in
Eq.(\ref{gr-kot10}), the most general expression of this type appears
to be
\begin{eqnarray}\label{gr-kot11}
  \chi^{\b\g}{}_{\a}{}^{\nu\rho}{}_\mu &=& {\frac 2\varkappa}\Big\{
{\b_1}g_{\a\m} 
\left(g^{\b\n}g^{\g\r}-g^{\c\n}g^{\b\r}\right)+\nonumber\\
&&{\b_2}\left[\left(g^{\c\r}\delta^\b_\a- g^{\b\r}\delta^\c_\a\right)\delta_\m^\n-
\left(g^{\c\n}\delta^\b_\a-g^{\b\n}\delta^\c_\a\right)\delta_\m^\r
                                          \right]+\nonumber\\
&&{\b_3}\left[\left(g^{\g\rho}\delta_\mu^\b -g^{\b\rho}\delta_\mu^\g\right)
\delta^\nu_\a -\left(g^{\g\nu}\delta_\mu^\b + g^{\b\nu}\delta_\mu^\g \right)
\delta^\rho_\a\right]\Big\}\,,
\end{eqnarray}
provided we assume the additional ``paircom'' symmetry
\begin{equation}\label{gr-kot11a}
\chi^{\b\c}{}_{\a}{}^{\n\r}{}_\m =\chi^{\n\r}{}_\m {}^{\b\c}{}_{\a}\,.
\end{equation}
Here {$\b_1,\b_2,\b_3$} are dimensionless factors,
$\varkappa$ is a {dimensionful} constant.

A remark is in order concerning the dimensions. The coframe and the
gravitational field strength have the dimensions of a length,
$[\vartheta^\alpha] = [d\vartheta^\alpha] = \ell$.  Analogously, the
gravitational current and the gravitational excitation have the same
dimension as a momentum:
$[\Sigma_\alpha] = [H_\alpha] = \text{[momentum]} = {\frac
  {m\,\ell}{t}}=ft$.
As a result, $[F^\alpha\wedge H_\alpha] = ft\ell = [\text{action}]$.
Hence the Lagrangian has, indeed, the correct dimension of an
action. Consequently, the dimension of the constant $\varkappa$ is
obtained as the ratio of the dimension of $F^\alpha$ divided by the
dimension of $H_\alpha$, that is, we have
$[\varkappa] = {\frac {t}{m}}$. Thus,
$[\kappa] = [\frac{\varkappa}{c}]={\frac {t^2}{m\ell}}=\frac{1}{f}$ is
Einstein's gravitational constant. This demonstrates a remarkable
consistency of teleparallel gravity with Einstein's GR.

Observe that the symmetry (\ref{gr-kot11a}) allows the coframe model
to be derived from a Lagrangian.  Using the constitutive tensor
(\ref{gr-kot11}), we can write the coframe Lagrangian in (\ref{lag1}) as
\begin{eqnarray}
\hspace{-20pt}{}^{(\vartheta)}\Lambda&=&\frac 12 F^\a\wedge H_\a
=\frac{1}{4\varkappa} 
F_{\b\g\a} \left(\b_1F^{\b\g\a}+\b_2g^{\a\b}F_\nu{}^{\g\nu}
  +\b_3F^{\a\g\b}\right){\rm vol}\,\nonumber \\ &=&
{\frac 1{2\varkappa} 
F_{\b\g\a} \left(\a_1{}^{(1)}\!F^{\b\g\a}+\a_2{}^{(2)}\!F^{\b\g\a}+\a_3
{}^{(3)}\!F^{\b\g\a}\right){\rm vol}\,,}
\label{cof-lag}
\end{eqnarray}
{where ${}^{(I)}\!F^{\b\g\a}$ are the 3 irreducible
  pieces of the field strength, see \cite{Hehl:1994ue}.}

\subsection{GR in terms of coframe variables}

We constructed a set of coframe models parametrized by dimensionless
numerical parameters $\a_1,\a_2,$ and $\a_3$ that turns out to be very
similar to the electrodynamics system.  The question is: How {are}
these models connected to gravity, in particular to GR?

Recall that Einsteins theory is expressed by the field equation
 \begin{equation}\label{gr1}
R_{ij}-\frac 12 Rg_{ij}=\kappa T_{ij}\,.
\end{equation}
Here $\kappa = 8\pi G/c^4$, with Newton's gravitational constant $G$.
When the metric tensor (\ref{gr-kot8}) is substituted into the
left-hand side of (\ref{gr1}), we obtain an expression that includes
second order derivatives of the coframe components plus the product of
their first order derivatives. Exactly the same type of expressions we
have in the coframe field equation $dH_\a=\Sigma_\a$. Thus, for some
special values of the parameters $\alpha_1,\alpha_2,$ and $\alpha_3$,
we can recover standard GR from the coframe field equation.  

It seems technically simpler to deal with the Lagrangian.  Recall that
the left-hand side of (\ref{gr1}) is derived from the action
functional
 \begin{equation}\label{gr2}
{\cal W}= {\frac 1{2\kappa c}}\int R\sqrt{-g}\,d^4x\,.
\end{equation}
As it is well known, the scalar curvature $R$ and, in turn,
the Lagrangian in (\ref{gr2}) can be expressed as a sum of two parts:
a term that is quadratic in the first order derivatives of the metric
plus a total divergence. In particular, up to a total derivative,
Eq.(\ref{gr2}) can be represented, see \cite{RMP} Eq.(3.20), as
 \begin{equation}\label{gr2a}
{\cal W} = {\frac 1{2\kappa c}}\int g^{ij}\left(\Gamma_{li}{}^k\Gamma_{kj}{}^l
  -\Gamma_{lk}{}^k\Gamma_{ij}{}^l\right)\sqrt{-g}\,d^4x \,.
\end{equation}
The expression of this Lagrangian in terms of the coframe is
well-known. In a compact form, see \cite{Itin:2001bp}, this {\it
  teleparallel equivalent of GR} reads
 \begin{equation}\label{gr2b}
{\cal W}= {\frac 12}\int F^\a\wedge H_\a\,,
\end{equation}
where 
 \begin{eqnarray}\label{gr2c}
  H_\a &=& {\frac 1{\kappa c}}\,^{\star} \big[g_{\a\b}F^\b-g_{\a\b}\vt^\b
     \wedge(e_\g\rfloor F^\g)-2g_{\b\g}e_\a\rfloor(\vt^\b\wedge 
       F^\g)\big]\nonumber\\
 &=& \frac {1}{\kappa c}\,g_{\a\b}\,^{\star} \big(
-{}^{(1)}\!F^\b +2\,^{(2)}\!F^\b +\frac{1}{2}{}^{(3)}\!F^\b\big)\,.
\end{eqnarray}
In tensor form (\ref{gr2c}) can be found in \cite{Hehl:1980}, see
Eq.(A.15).

There is a long development of this teleparallel theory of
gravity. Relevant papers are, amongst others, Pellegrini \& Plebanski
\cite{Pellegrini}, Kaempffer \cite{Kaempfer}, Cho \cite{Cho}, Hehl,
Nitsch \& von der Heyde \cite{Hehl:1980}, Nitsch et al.\
\cite{Nitsch:1980}, Muench et al.\, \cite{Muench:1998ay}, Nester et
al.\ \cite{Tung:1998dw,Nester:1998mp,So:2008kr}, Obukhov \& Pereira
\cite{Obukhov:2002tm}, Itin
\cite{Itin:1999wi,Itin:2001xz,Itin:2003jp,Itin:2006pd, Itin:2013gva},
Maluf \cite{Maluf:2013gaa}, Aldrovandi \& Pereira
\cite{Aldrovandi:2013}. A review was given in
\cite{Blagojevic:2013xpa}.

We substitute (\ref{gr2c}) into (\ref{gr2b}) and
compare the result with the coframe Lagrangian (\ref{cof-lag}).
The values of the free parameters turn out to be 
 \begin{equation}\label{gr3}
\big(\b_1=1\,,\b_2=-4\,, \b_3=2\big)\,\qquad\text{and}\qquad
\big(\a_1=-1\,,\a_2=2\,,\a_3=\frac 1 2\big)\,.
\end{equation} 
Since (\ref{cof-action}) includes all possible Lagrangians that are
quadratic in the first order derivatives of the coframe components, we
found that the Hilbert-Einstein Lagrangian is a special case of a coframe
Lagrangian.

\subsection{Local coframe transformations}

In a premetric teleparallel formalism, GR turns out to be a special
case of a general coframe model with the specific parameters of
(\ref{gr3}).  This case, however, is very distinguished. Indeed,
standard GR and its teleparallel equivalent are invariant under {\it
  local} Lorentz transformations of the coframe field,
\begin{equation}\label{gr-kot12}
  \vartheta^{\a'}=L_{\a}{}^{\a'\!}\!(x)\,\vartheta^{\a}\,.
\end{equation}
It can be checked, see \cite{Cho}, that there exists, up to an
arbitrary multiplicative constant, only one set of free parameters
$(\a_1,\a_2,\a_3)$, which constitutes a locally Lorentz invariant
coframe model with invariant Lagrangian and field equation. Other
ingredients of the coframe model, such as field strength, excitation,
energy-momentum current, and Lorentz-type force, are not locally
invariant. This fact is very well known in GR, where the
energy-momentum of gravity cannot be defined in a covariant way.

\section{Lorentz force and geodesics}

Equations of motion for test particles in an external gravitational field
should not be postulated, they are rather the consequence of the conservation laws.
Most conveniently, one can derive the equations of motion with the help of
the multipole expansion methods by integrating conservation laws over an
extended test body. Here we confine our attention to the lowest (monopole)
order and consider the relativistic version of Newton's equation of motion 
with the gravitational Lorentz force on the right-hand side. We demonstrate 
that one can rewrite the latter as the standard geodesic equation of GR, 
provided we assume the metric dependent constitutive relation between the
momentum $p_\a$ and the velocity $u^\a$ of a test particle.

\subsection{Premetric equation of particle motion}

In accordance with the expression (\ref{cof-lor1d}) of the twisted
covector-valued 4-form for the Lorentz force, the equation of motion
of test particle reads, in the monopole approximation,
\begin{equation}\label{lor-force2}
\frac {dp_\a}{ds}=u^\b p_\g F_{\a\b}{}^\g \,.
\end{equation}
Here $p_\a$ is the integrated momentum (1st moment) of an extended body. 
The body is characterized by an infinite set of multipole moments which
are derived by integrating the energy-momentum current density $\Sigma_a{}^\b$ 
over a cross-section of body's world tube. In the lowest approximation, 
we neglect effects of the dipole and higher order moments \cite{note}. 

The equation (\ref{lor-force2}) is invariant under arbitrary smooth 
reparametrization of the curve $s\longrightarrow  \lambda(s)$. Thus, 
even being expressed via the length parameter $s$, equation of motion 
(\ref{lor-force2}) is premetric, provided we consider the momentum $p_\a$ 
and the 4-velocity $u^\a$ as independent variables. Moreover, 
Eq.~(\ref{lor-force2}) is invariant under a rescaling of the momentum 
$p_\a\longrightarrow  Cp_a$. This symmetry manifests Einstein's principle 
of equivalence of inertial and gravitational mass, which is valid 
even in the premetric framework.

Let us rewrite Eq.(\ref{lor-force2}) in a coordinate basis. Multiplying both 
sides of this equation by $\vt_i{}^\a$, we find
\begin{eqnarray}
\vt_i{}^\a\,\frac {d p_\a}{ds} &=& u^j \left(F_{\a\b}{}^\g
  \vt_i{}^\a\vt_j{}^\b\right)  p_\g \nonumber\\
&=& u^j\left(\partial_i\vt_{j}{}^\g 
- \partial_j\vt_{i}{}^\g\right)  p_\g\,.\label{lor-force5}
\end{eqnarray}
Consequently, 
\begin{equation}\label{lor-force9}
\vt_i{}^\a\frac {d p_\a}{ds}+\frac {d\vt_i{}^\a}{ds} p_\a
=u^j p_\g {\partial_i \vt_j{}^\g}\,.
\end{equation}
Thus, the equation of motion of a test particle takes the form 
\begin{equation}\label{lor-force10}
\frac {d p_i}{ds}=u^j p_\a  {\partial_i \vt_j{}^\a}\,. 
\end{equation}
This equation is metric-free, and it is valid in a general geometric 
background.

\subsection{Geodesic equation}

Eventually, the metric $g$ is introduced on the spacetime manifold. Recall 
the two equivalent representations of the metric tensor in terms of a
coframe $\vt^\a$ or of coordinates $x^i$, respectively:
\begin{equation}\label{lor-force11}
 g=g_{\a\b}\,\vt^\a\otimes\vt^\b=g_{ij}\,dx^i\otimes dx^j\,.
\end{equation}
We observe
\begin{equation}\label{lor-force12}
  p_\g {\partial_i \vt_j{}^\g}=  g_{\b\g}p^k\vt_k^\b{\partial_i \vt_j{}^\g}
  =  \frac 12 p^k {\partial_i g_{jk}}\,.
\end{equation}
As a result, (\ref{lor-force10}) is recast into
\begin{equation}\label{lor-force13}
\frac {dp_i}{ds}=  u^j {\partial_j p_i}=\frac 12  {\partial_i g_{jk}}p^j u^k\,. 
\end{equation}
 
So far, this equation contains two unknowns, the covector $p_i$, the
momentum, and the vector $u^i$, the velocity.  We now assume the {\it
  constitutive relation} between the momentum and the velocity of the
particle to be local and linear,
\begin{equation}\label{lor-force14x}
  p_i=m g_{ij}u^j\,,
\end{equation}
where $m$ is the mass of the particle. As a consequence,
(\ref{lor-force13}) reduces to
\begin{equation}\label{lor-force15}
\frac {du_i}{ds}=  u^j{\partial_j u_i}=\frac 12 {\partial_i g_{jk}}u^k u^j\,. 
\end{equation}
This is equivalent to the standard geodesic equation, see \cite{LL}:
\begin{equation}\label{lor-force16}
  \frac {d u^i}{ds}+\Gamma_{jk}{}^iu^j u^k =0\,.
\end{equation}

\subsection{Particle motion in an electromagnetic field}

The premetric framework above, which correctly produces a geodesic,
can be extended to an electric point charge. The total force should be
the sum of the gravitational and the electromagnetic Lorentz terms
\begin{eqnarray}
\frac {dp_\a}{ds} = u^\b p_\g F_{\a\b}{}^\g+qu^\b F_{\a\b}\,.\label{lor-force17}
\end{eqnarray}
Here $q$ is the lowest multipole moment arising from the integration of 
the electric current vector density $J^\a$ over a cross-section of body's 
world tube; it is interpreted as a total electric charge of a test body.
Using the constitutive relation (\ref{lor-force14x}), we then end up
with the standard equation of motion of a charge in a curved spacetime:
\begin{equation}\label{lor-force19}
  \frac {d u^i}{d s}+\Gamma_{jk}{}^iu^j u^k = {\frac qm}\,F^{ij}u_j\,.
\end{equation}

\section{Discussion}

\subsubsection*{A gauge view at gravity}

A gauge-theoretical understanding of gravitational theory was our
tool for arriving at a premetric version of general relativity, namely
teleparallelism, here specifically by considering a gauge theory of
the {\it translation group.} However, it is the semidirect product of
the translation group $T(4)$ with the {\it Lorentz group} $SO(1,3)$,
the Poincar\'e group $T(4)\rtimes SO(1,3)$, which is the group of
motion in Minkowski spacetime. The Poincar\'e group is connected with
the energy-momentum and spin angular momentum of matter as Noether
currents.

The gauging, that is, the localization of the Poincar\'e group, yields
the Poincar\'e gauge theory of gravity (PG), see the review
\cite{Blagojevic:2013xpa}, Part B. If the spin of matter is
suppressed, a (In\"on\"u-Wigner type) group contraction of the PG
leads to a translation gauge theory. This contraction is
mathematically very delicate and is conventionally done in a heuristic
manner. In this way, the teleparallelism theory is emerging. At the
same time it becomes intelligible why teleparallelism has a number of
unexpected and somewhat strange features. After all, the vanishing of
the curvature, that is, the defining characteristics of teleparallelism
theory, is hard to digest from a purely Einsteinian GR point of view
(as already Pauli remarked to Einstein in the 1920s). However, from
the point of view of PG, this is self-evident, since the curvature is
the gauge field strength of the Lorentz group---and the suppression of
the material spin, in turn, suppresses the Lorentz group as gauge
group. And thus the Pauli objection can be invalidated. By the same
token we recognize that teleparallelism can only be really understood
in the context of PG. It is not comprehensible as a stand-alone
theory.

\subsubsection*{Nonlocal extension of teleparallelism}

A further success of the gauge-theoretical view at GR can be listed:
When, in the early 2000s, Mashhoon recognized that Einstein's clock
hypothesis is not sustainable as soon as high translational and
rotational accelerations occur. Therefore, he looked for a classical
nonlocal extension of GR and of the Einstein field equation. In spite
of several attempts, he was not able to implement it on the basis of
the Einstein equation and GR.

Again, as soon as one looked at gravity from a gauge-theoretical
perspective, it evident of how one has to proceed: Switch from GR to
the teleparallel approach to gravity. Its structure is closely related
to electromagnetism. And in electromagnetism it is straightforward to
generalize a local and linear constitutive law to a {\it non\/}local
and linear framework---already Volterra pointed this out in the 1910s.

Mashhoon and one of the present authors \cite{Hehl:2008eu,Hehl:2009es}
took their ``teleparallel'' glasses and looked at the field equation
of gravity. Following Volterra, they set up a nonlocal framework for a
classical theory of gravity, extending thereby GR to the domain of high
accelerations. This nonlocal theory of gravity was worked out in some
detail by Mashhoon and collaborators and can be found in the
forthcoming monograph of Mashhoon \cite{Mashhoon:2017}. Quite
unexpectedly, nonlocal gravity is able to describe the cosmos without
taking recourse to dark matter, see the title of
\cite{Hehl:2008eu}. The nonlocal theory explains dark matter
straightforwardly. Up to now, the astrophysical data seem to speak in
favor of this new framework.

\subsubsection*{$U(1)$-axion field versus axial torsion vector field}

Consider axion electrodynamics \cite{Wilczek:1987mv}: The
$U(1)$-axion $a$ is present in the 3rd irreducible piece of the
electromagnetic constitutive tensor in (\ref{el-const}):
\begin{equation}\label{axion}
^{(3)}\chi^{\a\b\g\d}=a\epsilon^{\a\b\g\d}\,,\qquad
[{}^{(3)}\chi^{\a\b\g\d}]=1/(\text{electric resistance})\,.
\end{equation}
Similarly, the axial torsion piece
${\cal A}:=g_{\a\b}\,^{\star\!}(\vt^\a\wedge F^\b)$ is manifest in the 3rd
piece of the gravitational constitutive tensor in (\ref{gr-kot9}):
\begin{equation}\label{axitorsion}
^{(3)}\chi^{\b\g}{}_{\a}{}^{\nu\rho}{}_{\mu}\,,\qquad
[^{(3)}\chi^{\b\g}{}_{\a}{}^{\nu\rho}{}_{\mu}]=\text{mass}/\text{time}
=\text{force}/\text{velocity}=[1/\kappa c]\,.
\end{equation}
The explicit form of $^{(3)}\chi^{\b\g}{}_{\a}{}^{\nu\rho}{}_{\mu}$
can be read off most conveniently from the Lagrangian
(\ref{cof-lag}). Both quantities, the electrodynamical axion and
the axial torsion, should contribute to the axial anomaly of quantum field
theory, see Obukhov \cite{Obukhov:1982da}.

Moreover, Mielke et al.\ \cite{Mielke:2006zp} tentatively assumed that the axial
torsion ${\cal A}$, which is a geometric quantity characterizing
spacetime, can be chosen as the gradient of a pseudoscalar field
$\cal P$, that is, ${\cal A} =d {\cal P}$.  Subsequently, without any
physical argument to support it and without an appropriate dimensional
analysis, $\cal P$ is identified with the axion field $a$ of the
internal $U(1)$-symmetry of Peccei-Quinn. This is what we call an ad
hoc assumption. Moreover, our dimensional analysis in Eqs.(\ref{axion})
and (\ref{axitorsion}) above shows how far-fetched such an assumption
is. 

Similar attempts were made by Castillo-Felisola et al.\
\cite{Castillo-Felisola:2015ema}. Corral argued that they don't
consider torsion as a field strength related to translational gauging,
but rather that they rely on ``the geometrical interpretation of {\it
  torsion}.'' And this would make a difference. We cannot share this
optimism: What else than a geometric quantity is a translational gauge
field strength, after all?

One could try the ansatz, with the superscript ${}^{(\vt)}$ denoting the
constitutive tensor density for the coframe Lagrangian (\ref{cof-lag}),  
\begin{equation}\label{concl1}
    \widehat{\chi}^{\b\g}{}_{\a}{}^{\nu\rho}{}_\mu
    ={}^{(\vt)}\chi^{\b\g}{}_{\a}{}^{\nu\rho}{}_\mu +a' \ep^{\b\g\nu\rho}g_{\a\mu}
\end{equation}
in order to link (\ref{axion}) with (\ref{axitorsion}). However, the
trace via $g^{\a\mu}$ of (\ref{concl1}) can never yield the axion,
unless one introduces in an ad hoc fashion a dimensionful factor in
$a'$. In other words, in this way one cannot find an axion in a
natural way.

The $U(1)$-axion is related to the {\it internal} group $U(1)$,
whereas the axial torsion is related to the {\it external} translation
group $T(4)$ via the Cartan circuit. One should not marry internal and
external groups, unless one investigates supersymmetry, which allows
such a mixing under certain circumstances.
\bigskip

\noindent{\bf Acknowledgments:} We are grateful to Crist\'obal Corral
and Oscar Castillo-Felisola (both from Valpara\'{\i}so, Chile) and to
Eckehard Mielke (Mexico City) for explaining to us their papers. For YNO
this work was partially supported by the Russian Foundation for Basic 
Research (Grant No. 16-02-00844-A).

\begin{appendix}

\section{Variation of the coframe Lagrangian}

We start with the premetric coframe Lagrangian
\begin{equation}\label{ap1}
  \Lambda=\frac 12 F^\a\wedge H_\a(F^\b)\,.
\end{equation}
Substituting the components of the forms
(\ref{gr-kot2a}), (\ref{gr-kot5}), we obtain
 \begin{equation}\label{ap2}
   \Lambda=\frac 18 \left(F_{\b\c}{}^\a
     \check{H}^{\m\n}{}_\a\right)\vt^{\b\c}\wedge \epsilon_{\m\n}\,.
\end{equation}
Applying the relation, which is a direct consequence of (\ref{vev}),
 \begin{equation}\label{ap3}
\vt^{\b\c}\wedge \epsilon_{\m\n}=\left(\d^\c_\m\d^\b_\n-\d^\b_\m\d^\c_\n\right)
{\rm vol}\,,
\end{equation}
we derive the coframe Lagrangian in components,
 \begin{equation}\label{ap4}
\Lambda=\frac 14 \left(F_{\b\c}{}^\a \check{H}^{\b\c}{}_\a\right){\rm vol}\,.
\end{equation}
Consequently the variation of the Lagrangian takes the form
 \begin{equation}\label{ap5}
   \d\Lambda=\frac 14 \left[\d \left(F_{\b\c}{}^\a\right)\check{H}^{\b\c}{}_\a
     + F_{\b\c}{}^\a\d  \left(\check{H}^{\b\c}{}_\a\right)\right]{\rm
     vol}+\frac 14  
   \left(F_{\b\c}{}^\a \check{H}^{\b\c}{}_\a\right)\d \left({\rm vol}\right)\,.
\end{equation}
Applying the local and linear constitutive relation (\ref{gr-kot9})
together with its symmetry property (\ref{gr-kot11a}), we find
\begin{eqnarray}\label{ap6}
F_{\b\c}{}^\a\d \left(\check{H}^{\b\c}{}_\a\right)&=&F_{\b\c}{}^\a
 \chi^{\b\c}{}_\a{}^{\n\r}{}_\m\d\left(F_{\n\r}{}^\m\right)\nonumber\\
&=&\d\left( F_{\n\r}{}^\m\right) \chi^{\n\r}{}_\m{}^{\b\c}{}_\a F_{\b\c}{}^\a
=\d\left( F_{\n\r}{}^\m\right) \check{H}^{\n\r}{}_\m\,.
\end{eqnarray}
Thus, Eq.(\ref{ap5}) takes the form
 \begin{equation}\label{ap7}
   \d\Lambda=\frac 12\d \left( F_{\b\c}{}^\a\right) \check{H}^{\b\c}{}_\a{\rm vol}
   +\frac 14  F_{\b\c}{}^\a \check{H}^{\b\c}{}_\a\,\d\left( {\rm vol}\right)\,.
\end{equation}
In order to calculate the variation $\d\left(F_{\b\c}{}^\a\right)$, we
use (\ref{gr-kot5}):
 \begin{equation}\label{ap8}
   \d (d\vt^\a)=\frac 12 \d\left(F_{\b\c}{}^\a\right)\vt^{\b\c}
   + F_{\b\c}{}^\a\d\vt^\b\wedge\vt^\c\,.
\end{equation}
Hence, 
 \begin{equation}\label{ap9}
   \d\left(F_{\b\c}{}^\a\right)=e_\c\rfloor e_\b\rfloor d(\d\vt^\a)
   -F_{\b\m}{}^\a e_\c\rfloor (\d\vt^\m)+F_{\c\m}{}^\a e_\b\rfloor (\d\vt^\m)\,.
\end{equation}
Thus, the first term of (\ref{ap7}) reads 
\begin{equation}\label{ap10}
  \frac 12\d\left(F_{\b\c}{}^\a\right)\check{H}^{\b\c}{}_\a{\rm vol}=-d(\d\vt^\a)
  \wedge H_\a-F_{\b\m}{}^\a \check{H}^{\b\c}{}_\a\d\vt^\m\wedge(e_\c
  \rfloor{\rm vol})\,.
\end{equation}
In order to calculate the variation of the volume element, we apply
the formula
 \begin{equation}\label{ap11}
\d({\rm vol})=\d\vt^\m\wedge(e_\m\rfloor{\rm vol})\,.
\end{equation}
Accordingly, the variation of the coframe Lagrangian (\ref{ap7}) takes
the form
\begin{equation}\label{ap12}
\d\Lambda=-d(\d\vt^\a)\wedge H_\a - \Sigma_\a\wedge \d\vt^\a\,,
\end{equation}
where 
\begin{equation}\label{ap13}
\Sigma_\a=\left(F_{\b\a}{}^\m \check{H}^{\b\r}{}_\m-\frac 14 \d^\r_\a
  F_{\b\c}{}^\m \check{H}^{\b\c}{}_\m\right)(e_\r\rfloor{\rm vol})\,.
\end{equation}
Using the components of the forms (\ref{gr-kot2a}), (\ref{gr-kot5}), we
obtain (\ref{cof-energy}) and (\ref{cof-energy1}). We extract the
total derivative as in (\ref{ap12}) and obtain finally
\begin{equation}\label{ap14}
\d\Lambda=-d(\d\vt^\a\wedge H_\a) -\d\vt^\a\wedge( d H_\a- \Sigma_\a)\,.
\end{equation}

\end{appendix}

\end{document}